\documentclass[twocolumn]{aastex62}

\usepackage{mathptmx}
\usepackage{verbatim}
\usepackage{graphicx}
\usepackage{amsmath}
\usepackage{textcomp}
\usepackage{latexsym}
\usepackage{multirow}
\usepackage{wasysym}
\usepackage{natbib}
\usepackage{subfigure}
\usepackage{booktabs}
\usepackage{longtable}
\usepackage{float}
\usepackage{tablefootnote}
\restylefloat{table}

\newcommand{\gtsim}{\raisebox{-.5ex}{$\;\stackrel{>}{\sim}\;$}}
\newcommand{\kms}{\ifmmode {\rm km\ s}^{-1} \else km s$^{-1}$\fi}

\newcommand{\lledd}{$L/L_{\rm Edd}$}

\newcommand{\et}{et al.\ }
\newcommand{\xray}{\hbox{X-ray}}
\newcommand{\aox}{$\alpha_{\rm ox}$}

\newcommand{\daox}{$\Delta\alpha_{\rm ox}$}
\newcommand{\nh}{$N_{\rm H}$}

\newcommand{\chandra}{{\sl Chandra}}

\newcommand{\hb}{H$\beta$}
\newcommand{\civ}{C~{\sc iv}}
\newcommand{\mgii}{Mg~{\sc ii}}

\journalinfo{The Astrophysical Journal, ???:??? (??pp), 2022 ??? ??}
\received{2021 September 17} 
\accepted{2022 March 21}

\shortauthors{MARLAR ET AL.}

\begin{document}
\title{Can X-ray Observations Improve Optical-UV-based Accretion-Rate Estimates for Quasars?}
\author{Andrea~Marlar}
\affiliation{
	Department of Physics, University of North Texas, Denton, TX 76203, USA; Andrea.Marlar@unt.edu} 
\author{Ohad~Shemmer}
\affiliation{
	Department of Physics, University of North Texas, Denton, TX 76203, USA; Andrea.Marlar@unt.edu} 
\author{Michael~S.~Brotherton}
\affiliation{
	Department of Physics and Astronomy, University of Wyoming, Laramie, WY 82071, USA} 
\author{Gordon~T.~Richards}
\affiliation{
	Department of Physics, Drexel University, 3141 Chestnut Street, Philadelphia, PA 19104, USA}
\author{Cooper~Dix}
\affiliation{
	Department of Physics, University of North Texas, Denton, TX 76203, USA; Andrea.Marlar@unt.edu}

\begin{abstract}
Current estimates of the normalized accretion rates of quasars (\lledd), rely on measuring the velocity widths of broad optical-UV emission lines (e.g., \hb\ and \mgii~$\lambda2800$). However, such lines tend to be weak or inaccessible in the most distant quasars, leading to increasing uncertainty in \lledd\ estimates at $z > 6$. Utilizing a carefully selected sample of 53 radio-quiet quasars that have \hb\ and \civ~$\lambda1549$ spectroscopy as well as \chandra\ coverage, we searched for a robust accretion-rate indicator for quasars, particularly at the highest-accessible redshifts ($z \sim 6-7$). Our analysis explored relationships between the \hb-based \lledd, the equivalent width (EW) of \civ, and the optical-to-\xray\ spectral slope (\aox). Our results show that EW(\civ) is the strongest indicator of the \hb-based \lledd\ parameter, consistent with previous studies, although significant scatter persists particularly for sources with weak \civ\ lines. We do not find evidence for the \aox\ parameter improving this relation, and we do not find a significant correlation between \aox\ and \hb-based \lledd. This absence of an improved relationship may reveal a limitation in our sample. \xray\ observations of additional luminous sources, found at $z \gtrsim 1$, may allow us to mitigate the biases inherent in our archival sample and test whether \xray\ data could improve \lledd\ estimates. Furthermore, deeper \xray\ observations of our sources may provide accurate measurements of the hard-\xray\ power-law photon index ($\Gamma$), which is considered an unbiased \lledd\ indicator. Correlations between EW(\civ) and \aox\ with $\Gamma$-based \lledd\ may yield a more robust prediction of a quasar normalized accretion rate.

\end{abstract}
\section{Introduction}
\label{sec:introduction}
Understanding the rapid growth of supermassive black holes and the assembly of their host galaxies requires reliable estimates of the black-hole mass and accretion rate, particularly in distant quasars. Currently, the most prominent method for doing so is by estimating the Eddington ratio, \lledd, where $L$ is the bolometric luminosity and $L_{\rm{Edd}}$ is the Eddington luminosity. This method relies on measuring the velocity widths of prominent optical-UV broad emission lines (e.g., \hb\ and \mgii~$\lambda2800$) from single-epoch spectra, assuming that the broad emission line region is virialized (e.g., Laor 1998; Vestergaard \& Peterson 2006; Shen \& Liu 2012, Mej{\'\i}a-Restrepo \et 2016, Grier \et 2017, Du \et 2018, Bahk \et 2019, Dalla Bont{\`a} \et 2020). However, these lines tend to be weak or inaccessible in the most distant, and typically highly luminous, quasars, leading to increasingly uncertain accretion rates at high redshift (e.g., Ba{\~n}ados \et 2016, Onoue \et 2019, Reed \et 2019, Wang \et 2021). Therefore, there is a need for an alternative \lledd\ estimate that is reliable at $z \sim 6$ and beyond. 

One such estimate can be obtained from the hard-\xray\ power-law photon index ($\Gamma$; e.g., Shemmer \et 2006, 2008; Risaliti \et 2009; Brightman \et 2013), typically measured above a rest-frame energy of $2$ kev. However, the different rest-frame energies covered at different redshifts makes it difficult to measure this parameter accurately and, due to the larger exposure times necessary, it is currently not economical or practical to measure this parameter for a statistically meaningful number of distant sources (e.g., Moretti \et 2014; Page \et 2014; Nanni \et 2017). Therefore, the question remains whether a more practical \xray\ parameter could provide an equivalent, or improved, \lledd\ estimate.

Another indicator of \lledd\ is the equivalent width (EW) of the prominent \civ\ $\lambda1549$ emission line (e.g., Baskin \& Laor 2004; Shemmer \& Lieber 2015; Rivera \et 2020). However, there is significant scatter around this relationship, particularly due to weak emission-line quasars (WLQs) that lie systematically below the best fit EW(\civ)-\lledd\ relation (Shemmer \& Lieber 2015). 

The level of \xray\ emission from quasars with respect to their optical-UV emission is another possible diagnostic of their accretion rates. Previous studies have shown a strong correlation between the optical-\xray\ spectral slope, \aox\footnote{Defined as \aox$=\log(f_{\rm 2\,keV}/f_{2500\,\text{\AA}})/ \log(\nu_{\rm 2\,keV}/\nu_{2500\,\text{\AA}})$, where $f_{\rm 2\,keV}$ and $f_{2500\,\text{\AA}}$ are the flux densities at frequencies corresponding to $2$\,keV ($\nu_{\rm 2\,keV}$) and $2500$\,\AA\ ($\nu_{2500\,\text{\AA}}$), respectively.}, and luminosity (e.g., Just \et 2007, Lusso \et 2010, Timlin \et 2020). However, similarly strong or significant correlations between \aox\ and \lledd\ have not yet been found, presumably because \aox\ depends also on the black hole mass (e.g., Shemmer \et 2008, Grupe \et 2010, Wu \et 2012, Liu \et 2021). 

It is possible that some of the aforementioned studies have failed to find a significant correlation between \aox\ and \lledd\ due to a difficulty addressing this in a comprehensive manner that incorporates all the principal observable quantities. In this work we present an archival sample of quasars that have coverage from the \chandra\ \xray\ observatory\footnote{https://cxc.cfa.harvard.edu/csc/} (hereafter, \chandra; Weisskopf \et 2000), and have high-quality data in the \civ\ and \hb\ spectral bands. Our goal is to identify a combination of basic observable properties that can serve as a reliable and practical indicator of \lledd\ for quasars, particularly at ``Cosmic Dawn" ($z\gtsim6$). This approach allows us to jointly analyze all the principal diagnostics of the \lledd\ parameter, in spite of the fact that our sample is not statistically complete (see Figure~\ref{fig:contour}). We describe our sample selection, observations, and data reduction in Section 2; in Section 3 we present the results of our analyses. We summarize our findings in Section 4. Throughout this work we compute luminosity distances using the standard cosmological model (\hbox{$H_{0}$ = 70 km ${\rm s}^{-1}\,\ {\rm Mpc}^{-1}$}, \hbox{$\Omega_{\Lambda}$ = 0.7}, and \hbox{$\Omega_{M}$ = 0.3}; e.g., Spergel \et 2007). Complete source names appear in the Tables, and abbreviated names appear in Figures and throughout the text.

\section{Target Selection, Observations, and Data Reduction}
\label{sec:obs}
We selected sources from the Sloan Digital Sky Survey (SDSS) quasar catalog from Data Release 16 (DR16Q) (Lyke \et 2020), which was the largest, most uniform catalog of optically selected quasars at the time. We then narrowed the sample to sources that have high-quality optical spectra without broad absorption lines (BALs), and are radio quiet\footnote{The radio loudness parameter, $R$, is defined as $f_{\rm 5\,GHz}/f_{4400\,\text{\AA}}$, where $f_{\rm 5\,GHz}$ and $f_{4400\,\text{\AA}}$ are the flux densities at $5$\,GHz and $4400$\,\AA, respectively (Kellermann \et 1989).} (\hbox{$R<10$}). This step removed $\sim5\%$ of sources from the original catalog. Our sample is further limited to sources within the redshift ranges $0\lesssim z\lesssim 0.8$ and $1.5\lesssim z\lesssim 3.6$ (removing an additional $\sim30\%$ of sources from the original catalog). The former assures that the \hb\ region is covered by SDSS spectra, and \civ\ is covered by high-quality, rest-frame UV spectra, from the \textit{Hubble Space Telescope (HST)} except for two sources (SDSS J0057+1446 and SDSS J0159+0023) that were measured from \textit{International Ultraviolet Explorer} (IUE) spectra; most of the sources in this low-luminosity sub-sample have been selected as bright UV sources that have a relatively narrow range in UV luminosity as described in detail in Rivera \et (2022, under review).

\begin{figure}
\plotone{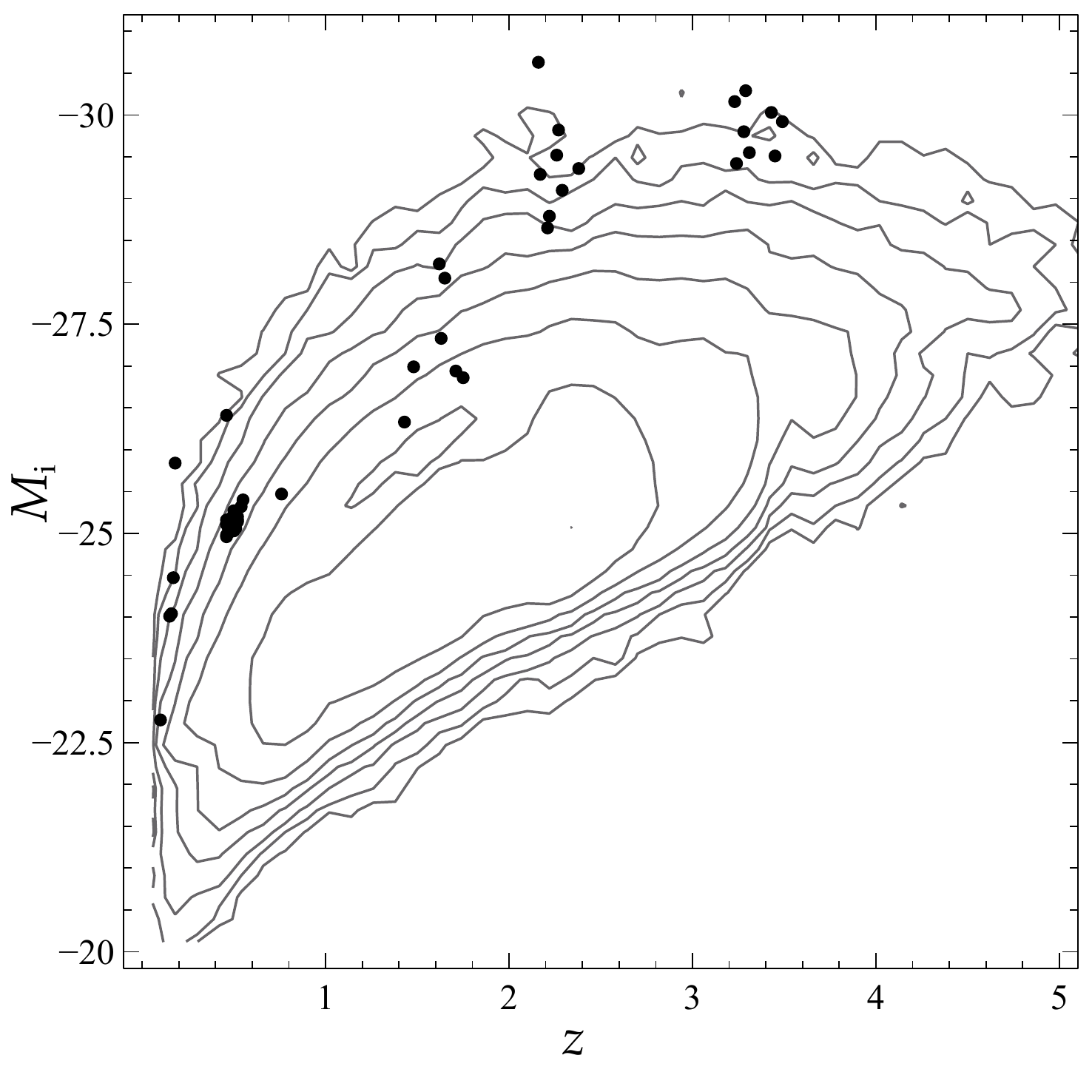}
\caption{The absolute \textit{i}-band magnitude vs. redshift for our sample (black dots) plotted against the SDSS DR16 quasar catalog sample (contours).} 
\label{fig:contour}
\end{figure}

 The $1.5\lesssim z\lesssim 3.6$ range is split into three narrower intervals, $1.5\lesssim z\lesssim 1.7$, $2.0\lesssim z\lesssim 2.5$, and $3.1\lesssim z\lesssim 3.6$ (removing an additional $\sim23\%$ of sources from the original catalog), which assures that the \hb\ line has near infrared (NIR) spectroscopic coverage in the $J$, $H$, and $K$ bands, respectively; the \civ\ line at these redshifts is covered by SDSS spectra. Figure~\ref{fig:contour} shows the luminosity vs. redshift distribution of our sample with respect to all SDSS DR16 quasars. It should be noted that, drawn from diverse archival samples for the non-SDSS data, our sources do not uniformly span the parameter space of interest; however, they can still be used to establish the analysis approach presented herein and identify regions of parameter space that require additional \xray\ observations.

\begin{figure}
\plotone{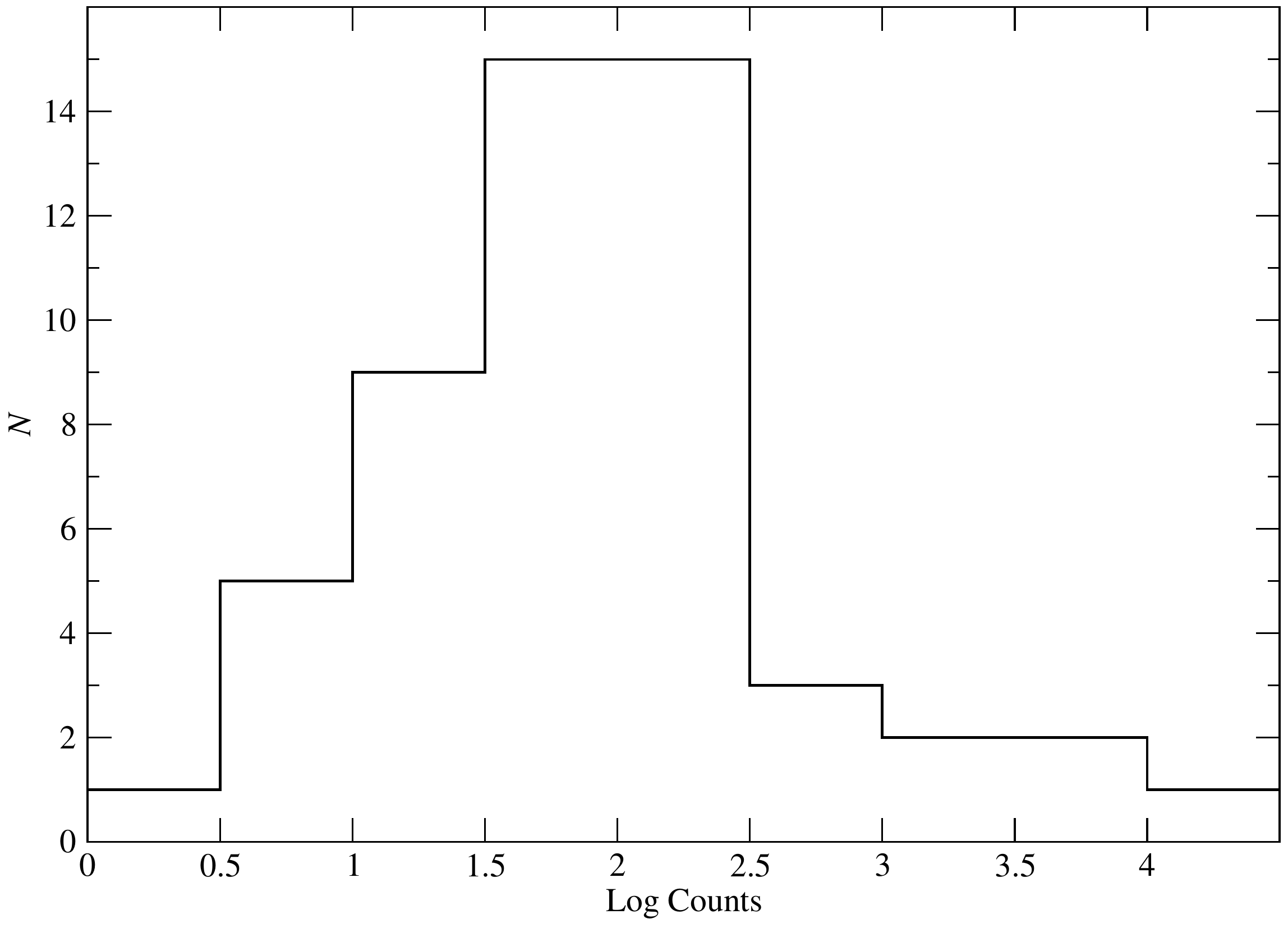}
\caption{Histogram of the number of counts for each source in the observed-frame $0.5-8$ keV band.} 
\label{fig:hist}
\end{figure}

We cross-matched the optical-UV sample with the \chandra\ archive for high-quality \xray\ imaging spectroscopy in the observed-frame $0.5-10$ keV energy range. To minimize spurious detections, we constrained our sample to objects with an optical-\xray\ angular distance ($\Delta_{\rm{Opt-X}}$; i.e., the positional offset between the SDSS and \chandra\ coordinates) of $< 1''$, and compiled a final sample of $53$ sources, about half of which are at $z\lesssim 0.8$. This seemingly small number is a consequence of starting off with about three quarters of a million SDSS quasars but only $< 1$\% of which are at low redshift and have high quality spectral coverage of the \civ\ region in ultraviolet spectra, and similarly, $< 1$\% of which are at high redshift and have high quality spectral coverage of the \hb\ region. All sources were targeted for \chandra\ observations, and all but one were observed with the \chandra\ Advanced CCD Imaging Spectrometer (ACIS; Garmire \et 2003); SDSS J1119$+$2119 was observed with the \chandra\ High Resolution Camera (HRC; Murray \et 1997). Most of the \chandra\ observations are considered ``snapshots" (i.e., $\lesssim 300$ counts), and $\sim15\%$ of the observations are considered as being ``deep" (as can be seen by the sharp drop above $\sim300$ counts in Figure~\ref{fig:hist}).

The \chandra\ observation log appears in \hbox{Table~\ref{tab:chandra_log}}. \textit{Column (1)} is the SDSS quasar name; \textit{Column (2)} is the redshift from the SDSS DR16 quasar catalog; \textit{Column (3)} gives the angular distance between the optical and \xray\ positions; \textit{Column (4)} shows the Galactic absorption column density in units of $10^{20} {\rm cm}^{-2}$, taken from Dickey $\&$ Lockman (1990) and obtained with the HEASARC \nh\ tool\footnote{https://heasarc.gsfc.nasa.gov/cgi-bin/Tools/w3nh/w3nh.pl.}; \textit{Columns (5) - (8)} give the \chandra\ Cycle, start date, observation ID, and exposure time, respectively. 

Source counts were extracted using \chandra\ Interactive Analysis of Observations ({\sc ciao})\footnote{http://cxc.cfa.harvard.edu/ciao/} v4.10 tools. The \xray\ counts for all sources except SDSS J1119$+$2119 were obtained using {\sc wavdetect} (Freeman \et 2002) with wavelet transforms of scale sizes $1$, $1.4$, $2$, $2.8$, and $4$ pixels, a false-positive probability threshold of $10^{-3}$, and confirmed by visual inspection. Source counts for SDSS J1119$+$2119 were estimated using the {\sc ciao} {\sc dmextract} script in the HRC wide band (observed-frame \hbox{$0.1-10$ keV}); the \chandra\ {\sc pimms}\footnote{https://cxc.harvard.edu/toolkit/pimms.jsp} v4.10 tool was then used to estimate the counts in the energy bands as described below.

\startlongtable
\begin{deluxetable*}{lcccclcc}
\tabletypesize{\footnotesize}
\tablecolumns{8}
\tablecaption{\chandra\ Observation Log \label{tab:chandra_log}}
\tablehead{
\colhead{} &
\colhead{} &
\colhead{$\Delta_{\rm{Opt-X}}$} &
\colhead{Galactic \nh} &
\colhead{} &
\colhead{} &
\colhead{} &
\colhead{Exp. Time\tablenotemark{a}} \\
\colhead{Quasar} &
\colhead{$z$} &
\colhead{(arcsec)} &
\colhead{(10$^{20}$\,cm$^{-2}$)} &
\colhead{Cycle} &
\colhead{Obs. Date} & 
\colhead{Obs. ID} &
\colhead{(ks)} \\
\colhead{(1)} &
\colhead{(2)} &
\colhead{(3)} &
\colhead{(4)} &
\colhead{(5)} &
\colhead{(6)} &
\colhead{(7)} &
\colhead{(8)} 
}
\startdata
\object{SDSS~J002019.22$-$110609.2} & 0.49 & 0.1 & 2.89 & 18 & 2017 Jan 16 & 19535 & 3.50 \\
\object{SDSS~J005709.94$+$144610.1} & 0.17 & 0.1 & 4.37 & 1 & 2000 Jul 28 & 865 & 4.66 \\
\object{SDSS~J014812.83$+$000322.9} & 1.48 & 0.2 & 2.89 & 9 & 2008 Nov 9 & 9225 & 10.54 \\
\object{SDSS~J015950.23$+$002340.9} & 0.16 & 0.8 & 2.59 & 4 & 2003 Aug 26 & 4104 & 9.73 \\
\object{SDSS~J030341.04$-$002321.9} & 3.23 & 0.3 & 7.14 & 13 & 2011 Nov 27 & 13349 & 1.54 \\
\object{SDSS~J032349.53$-$002949.8} & 1.63 & 0.2 & 6.71 & 6 & 2005 Oct 30 & 5654 & 8.31 \\
\object{SDSS~J080117.79$+$521034.5} & 3.21 & 0.6 & 4.66 & 15 & 2014 Dec 11 & 17081 & 43.50 \\
\object{SDSS~J082024.21$+$233450.4} & 0.47 & 0.2 & 4.02 & 18 & 2017 Feb 1 & 19536 & 2.95 \\
\object{SDSS~J082658.85$+$061142.6} & 0.50 & 0.4 & 2.68 & 18 & 2016 Dec 29 & 19537 & 3.43 \\
\object{SDSS~J083332.92$+$164411.0} & 0.46 & 0.3 & 3.60 & 18 & 2017 Oct 12 & 19538 & 2.95 \\
\object{SDSS~J083510.36$+$035901.1} & 0.49 & 0.2 & 3.29 & 18 & 2017 Jun 12 & 19539 & 2.95 \\
\object{SDSS~J084846.11$+$611234.6} & 2.26 & 0.1 & 4.43 & 13 & 2011 Dec 22 & 13353 & 1.54 \\
\object{SDSS~J085116.14$+$424328.8} & 0.48 & 0.3 & 2.56 & 18 & 2017 Jan 15 & 19540 & 2.95 \\
\object{SDSS~J090033.50$+$421547.0} & 3.29 & 0.1 & 1.99 & 7 & 2006 Feb 9 & 6810 & 3.91 \\
\object{SDSS~J091451.42$+$421957.0} & 0.55 & 0.3 & 1.46 & 18 & 2017 Jan 11 & 19541 & 3.51 \\  
\object{SDSS~J093502.52$+$433110.6} & 0.46 & 0.2 & 1.40 & 18 & 2017 Jan 12 & 19542 & 2.89 \\
\object{SDSS~J094202.04$+$042244.5} & 3.27 & 0.2 & 3.56 & 7 & 2006 Feb 8 & 6821 & 4.07 \\
\object{SDSS~J094602.31$+$274407.0} & 2.44 & 0.1 & 1.77 & 11 & 2010 Jan 16 & 11489 & 4.98 \\
\object{SDSS~J094646.94$+$392719.0} & 2.22 & 0.3 & 1.57 & 12 & 2011 Feb 27 & 12857 & 27.30 \\
\object{SDSS~J095852.19$+$120245.0} & 3.30 & 0.1 & 3.22 & 13 & 2012 Apr 22 & 13354 & 1.56 \\
\object{SDSS~J100054.96$+$262242.4} & 0.51 & 0.4 & 2.68 & 18 & 2017 Mar 4 & 19543 & 3.50 \\
\object{SDSS~J102907.09$+$651024.6} & 2.18 & 0.2 & 1.20 & 9 & 2008 Jun 17 & 9228 & 10.64 \\
\object{SDSS~J103320.65$+$274024.2} & 0.54 & 0.1 & 1.87 & 18 & 2017 Feb 1 & 19544 & 3.51 \\
\object{SDSS~J111119.10$+$133603.8} & 3.48 & 0.2 & 1.57 & 16 & 2015 Jan 26 & 17082 & 43.06 \\
\object{SDSS~J111138.66$+$575030.0} & 0.47 & 0.2 & 0.71 & 18 & 2017 Aug 31 & 19545 & 2.98 \\
\object{SDSS~J111830.28$+$402554.0} & 0.15 & $<0.1$ & 1.91 & 1 & 2000 Oct 3 & 868 & 19.70 \\
\object{SDSS~J111908.67$+$211918.0} & 0.18 & 0.3 & 1.28 & 3 & 2002 Jun 29 & 3145 & 88.05 \\
\object{SDSS~J111941.12$+$595108.7} & 0.49 & 0.2 & 0.73 & 18 & 2017 Aug 12 & 19546 & 3.54 \\
\object{SDSS~J112224.15$+$031802.6} & 0.47 & $<0.1$ & 4.16 & 18 & 2017 Jan 28 & 19547 & 2.95 \\
\object{SDSS~J112614.93$+$310146.6} & 0.49 & 0.1 & 1.76 & 18 & 2017 Jan 23 & 19548 & 3.50 \\
\object{SDSS~J113327.78$+$032719.1} & 0.52 & 0.7 & 2.74 & 18 & 2017 Jan 27 & 19549 & 3.43 \\
\object{SDSS~J115954.33$+$201921.1} & 3.43 & 0.7 & 2.39 & 13 & 2012 Feb 28 & 13317 & 1.56 \\
\object{SDSS~J123734.47$+$444731.7} & 0.46 & 0.1 & 1.50 & 18 & 2017 Mar 3 & 19551 & 2.95 \\
\object{SDSS~J125415.55$+$480850.6} & 0.50 & 0.5 & 1.12 & 18 & 2017 Apr 5 & 19552 & 3.05 \\
\object{SDSS~J131627.84$+$315825.7} & 0.46 & 0.4 & 1.11 & 18 & 2017 Jan 25 & 19553 & 3.43 \\
\object{SDSS~J134701.54$+$215401.1} & 0.50 & 0.3 & 1.63 & 18 & 2017 Mar 22 & 19554 & 3.50 \\
\object{SDSS~J135023.68$+$265243.1} & 1.62 & 0.3 & 1.23 & 16 & 2015 Apr 5 & 17225 & 58.76 \\
\object{SDSS~J140331.29$+$462804.8} & 0.46 & 0.4 & 1.26 & 18 & 2017 Apr 20 & 19555 & 2.95 \\
\object{SDSS~J140621.89$+$222346.5} & 0.10 & 0.3 & 2.14 & 1 & 2000 Jul 22 & 812 & 79.12 \\
\object{SDSS~J141028.14$+$135950.2} & 2.21 & 0.1 & 1.42 & 10 & 2009 Nov 28 & 10741 & 4.03 \\
\object{SDSS~J141141.96$+$140233.9} & 1.75 & 0.1 & 1.43 & 14 & 2012 Dec 16 & 15353 & 3.39 \\
\object{SDSS~J141730.92$+$073320.7} & 1.70 & 0.4 & 2.12 & 14 & 2012 Dec 5 & 15349 & 2.48 \\
\object{SDSS~J141949.39$+$060654.0} & 1.64 & $<0.1$ & 2.20 & 9 & 2008 Mar 8 & 9226 & 9.92 \\
\object{SDSS~J141951.84$+$470901.3} & 2.30 & $<0.1$ & 1.52 & 3 & 2002 Jun 2 & 3076 & 7.66 \\
\object{SDSS~J144741.76$-$020339.1} & 1.43 & 0.3 & 4.53 & 14 & 2013 Jan 13 & 15355 & 2.00 \\
\object{SDSS~J145334.13$+$311401.4} & 0.46 & 0.7 & 1.47 & 18 & 2017 Jan 31 & 19556 & 2.98 \\
\object{SDSS~J152156.48$+$520238.5} & 2.21 & 0.9 & 1.58 & 14 & 2013 Oct 22 & 15334 & 37.39 \\
\object{SDSS~J152654.61$+$565512.3} & 0.48 & 0.5 & 1.42 & 18 & 2017 Feb 13 & 19557 & 3.50 \\
\object{SDSS~J155837.77$+$081345.8} & 0.52 & $<0.1$ & 3.68 & 18 & 2017 Jan 21 & 19558 & 3.43 \\
\object{SDSS~J212329.46$-$005052.9} & 2.27 & 0.5 & 3.65 & 16 & 2015 Dec 22 & 17080 & 39.55 \\
\object{SDSS~J230301.45$-$093930.7} & 3.46 & 0.3 & 3.32 & 13 & 2011 Dec 24 & 13358 & 1.54 \\
\object{SDSS~J234145.51$-$004640.5} & 0.52 & 0.3 & 3.67 & 18 & 2017 Jun 22 & 19559 & 3.43 \\ 
\object{SDSS~J235321.62$-$002840.6} & 0.76 & 0.1 & 3.45 & 7 & 2006 May 12 & 6876 & 2.85 
\enddata
\tablecomments{\textit{Column (1)} is the SDSS quasar name; \textit{Column (2)} is the redshift from the SDSS DR16 quasar catalog; \textit{Column (3)} gives the angular distance between the optical and \xray\ positions; \textit{Column (4)} shows the Galactic absorption column density in units of $10^{20} {\rm cm}^{-2}$, taken from Dickey $\&$ Lockman (1990) and obtained with the HEASARC \nh\ tool; \textit{Columns (5) - (8)} give the \chandra\ Cycle, start date, observation ID, and exposure time, respectively.}
\tablenotetext{a}{The exposure time has been corrected for detector dead time.}
\end{deluxetable*}

Table 2 presents the basic \xray\ measurements and UV-optical data used for our analyses. \textit{Column (1)} is the SDSS quasar name; \textit{Columns (2) - (4)} give the \xray\ counts in the soft (observed-frame \hbox{$0.5-2$ keV}), hard (observed-frame \hbox{$2-8$ keV}), and full (observed-frame \hbox{$0.5-8$ keV}) bands, respectively; \textit{Column (5)} gives the count rate in the soft band; \textit{Column (6)} gives the Galactic absorption-corrected flux density at rest-frame $2$ keV, assuming a power-law model with $\Gamma=2.0$; \textit{Column (7)} gives the optical flux density at rest-frame $2500$\,\text{\AA}; \textit{Column (8)} is the \aox\ parameter; \textit{Column (9)} gives the $\Delta$\aox\ parameter, which is the difference between the measured \aox\ from Column (8) and the predicted \aox , based on the \hbox{\aox-$L_{\nu}(2500\,\text{\AA})$} relation in quasars (given as eq. [3] of Timlin \et 2020); \textit{Column (10)} gives the monochromatic luminosity at a rest-frame wavelength of $5100\,\text{\AA}$ [$\nu L_{\nu}(5100\,\text{\AA})$] computed from the flux densities in Column (7), assuming a UV-optical power-law slope of $\alpha_{\rm uv} = -0.5$ (e.g., Vanden Berk \et 2001); \textit{Columns (11) and (12)} are the archival measurements and respective references for the FWHM of the broad \hb\ line; \textit{Column (13)} is the Eddington ratio, derived using eq. [2] of Shemmer \et (2010),
\begin{equation}
L/L_{\rm Edd}=0.13 f(L) \left [ \frac{\nu L_{\nu}(5100\,\text{\AA})}{10^{44}{\rm~erg~s}^{-1}} \right ]^{0.5} \left [ \frac{{\rm FWHM(H}\beta {\rm )}}{10^3 {\rm~km~s}^{-1}} \right ]^{-2},
\end{equation}
where $f(L)$ is the luminosity-dependent bolometric correction to $\nu L_{\nu}(5100\,\text{\AA})$, and was computed using eq. [21] of Marconi \et (2004); \textit{Column (14)} gives the rest-frame \civ\ equivalent width as described below. 

The \civ\ emission line was fit with a local, linear continuum and two independent Gaussian profiles. The linear continuum was constructed using the rest-frame fitting windows $[1445,1465]\,\text{\AA}$ and $[1700,1705]\,\text{\AA}$. The Gaussians were constrained such that the line peak would lie within \hbox{$5,000$ km s$^{-1}$} from the wavelength that corresponded to the maximum flux density of the emission line, the widths could range from \hbox{$0$ km s$^{-1}$} to $15,000$ km s$^{-1}$, and the flux density was constrained to up to twice the maximum value of the emission line. Each fit was visually inspected to avoid narrow absorption lines within the \civ\ profile and noise spikes in the continuum fitting windows. We computed the EW(\civ) values for our sources and compared $21$ of them with the respective values that were available in Shen \et (2011). The difference between our values and those from Shen \et (2011), for non-WLQs, is non-systematic and ranges between $|0.8\%|$ and $|33.1\%|$ with a mean value of $|17.5\%|$. The values for the WLQs SDSS~J0946$+$2744, SDSS~J1411$+$1402, and SDSS~J1521$+$5202 differ from those of Shen \et (2011) by $|196.4\%|$, $|57.5\%|$, and $|204.1\%|$, respectively, which can be attributed to WLQs having extremely low EW(\civ) values with high uncertainties.

\newpage
\startlongtable
\movetabledown=0.25in
\begin{longrotatetable}
\begin{deluxetable*}{lccccccccccccc}
\tabletypesize{\footnotesize}
\tablecolumns{14}
\tablecaption{Basic X-ray and UV-Optical Data \label{tab:uv-opt}}
\tablehead{ 
\colhead{} &
\colhead{} &
\multicolumn{1}{c}{Counts\tablenotemark{a}} &
\colhead{} &
\colhead{} &
\colhead{} &
\colhead{} &
\colhead{} &
\colhead{} &
\colhead{$\log \nu L_{\nu}(5100\,\text{\AA})$} &
\colhead{FWHM \hb} &
\colhead{} &
\colhead{} &
\colhead{EW(\civ)} \\
\cline{2-4} 
\colhead{Quasar} &
\colhead{$0.5-2$~keV} & 
\colhead{$2-8$~keV} &
\colhead{$0.5-8$~keV} & 
\colhead{Count Rate} &
\colhead{$f_{2~\rm keV}$} &
\colhead{$f_{2500\,\text{\AA}}$} &
\colhead{\aox} &
\colhead{\daox} &
\colhead{(erg\,s$^{-1}$)} &
\colhead{(km\,s$^{-1}$)} &
\colhead{Ref.} &
\colhead{\lledd} &
\colhead{(\text{\AA})} \\
\colhead{(1)} &
\colhead{(2)} &
\colhead{(3)} &
\colhead{(4)} &
\colhead{(5)} &
\colhead{(6)} &
\colhead{(7)} &
\colhead{(8)} &
\colhead{(9)} &
\colhead{(10)} &
\colhead{(11)} &
\colhead{(12)} &
\colhead{(13)} &
\colhead{(14)}
}
\startdata
\object{SDSS~J002019.22$-$110609.2} & 36.5$^{+7.1}_{-6.0}$ & 21.9$^{+5.8}_{-4.6}$ & 58.3$^{+8.7}_{-7.6}$ & 10.4$^{+2.0}_{-1.7}$ & 1.6$\pm{0.3}$ & 2.4 & $-1.60$ & $-0.17$ & 45.1 & 3105.8 & 1 & 0.31 & 32.1 \\
\object{SDSS~J005709.94$+$144610.1} & 3793.5$^{+62.6}_{-61.6}$ & 843.9$^{+30.1}_{-29.0}$ & 4632.2$^{+69.1}_{-68.1}$ & 814.1$^{+13.4}_{13.2}$ & 45.9$\pm{0.7}$ & 23.5 & $-1.42$ & $+0.02$ & 45.1 & 10011.1 & 1 & 0.03 & 99.2 \\
\object{SDSS~J014812.83$+$000322.9} & 93.8$^{+10.7}_{-9.7}$ & 26.8$^{+6.2}_{-5.1}$ & 120.4$^{+12.0}_{-11.0}$ & 8.9$^{+1.0}_{-0.9}$ & 1.3$\pm{0.1}$ & 1.4 & $-1.55$ & $+0.02$ & 45.8 & 6475.0 & 2 & 0.15 & 15.4 \\
\object{SDSS~J015950.23$+$002340.9} & 3649.5$^{+61.4}_{-60.4}$ & 667.3$^{+26.8}_{-25.8}$ & 4316.9$^{+66.7}_{-65.7}$ & 375.2$^{+6.3}_{-6.2}$ & 21.3$\pm{0.4}$ & 24.6 & $-1.56$ & $-0.13$ & 45.1 & 2406.2 & 1 & 0.52 & 77.1 \\
\object{SDSS~J030341.04$-$002321.9} & 5.0$^{+3.4}_{-2.1}$ & 3.9$^{+3.2}_{-1.9}$ & 9.0$^{+4.1}_{-2.9}$ & 3.2$^{+2.2}_{-1.4}$ & 1.0$^{+0.7}_{-0.4}$ & 5.5 & $-1.82$ & $-0.01$ & 47.0 & 3010.0 & 3 & 2.51 & 41.0 \\
\object{SDSS~J032349.53$-$002949.8} & 131.8$^{+12.5}_{-11.5}$ & 32.7$^{+6.8}_{-5.7}$ & 164.4$^{+13.8}_{-12.8}$ & 15.9$^{+1.5}_{-1.4}$ & 3.0$\pm{0.3}$ & 2.2 & $-1.49$ & $+0.14$ & 46.1 & 2990.6 & 2 & 0.96 & 48.1 \\
\object{SDSS~J080117.79$+$521034.5} & 116.0$^{+11.8}_{-10.8}$ & 49.5$^{+8.1}_{-7.0}$ & 168.3$^{+14.0}_{-13.0}$ & 2.7$^{+0.3}_{-0.2}$ & 0.9$\pm{0.1}$ & 10.0 & $-1.94$ & $-0.08$ & 47.3 & 5448.7 & 4 & 1.07 & 13.8 \\
\object{SDSS~J082024.21$+$233450.4} & 96.9$^{+10.9}_{-9.8}$ & 26.4$^{+6.2}_{-5.1}$ & 122.9$^{+12.1}_{-11.1}$ & 32.8$^{+3.7}_{-3.3}$ & 5.2$^{+0.6}_{-0.5}$ & 3.7 & $-1.48$ & $-0.02$ & 45.2 & 2627.0 & 1 & 0.48 & 55.0 \\
\object{SDSS~J082658.85$+$061142.6} & 52.4$^{+8.3}_{-7.2}$ & 20.7$^{+5.6}_{-4.5}$ & 73.0$^{+9.6}_{-8.5}$ & 15.3$^{+2.4}_{-2.1}$ & 2.4$^{+0.4}_{-0.3}$ & 2.8 & $-1.56$ & $-0.11$ & 45.2 & 1941.0 & 1 & 0.88 & 25.5 \\
\object{SDSS~J083332.92$+$164411.0} & 73.5$^{+9.6}_{-8.6}$ & 37.6$^{+7.2}_{-6.1}$ & 111.6$^{+11.6}_{-10.5}$ & 24.9$^{+3.3}_{-2.9}$ & 3.9$\pm{0.5}$ & 3.4 & $-1.51$ & $-0.06$ & 45.2 & 2954.7 & 1 & 0.38 & 14.6 \\
\object{SDSS~J083510.36$+$035901.1} & 31.6$^{+6.7}_{-5.6}$ & 18.7$^{+5.4}_{-4.3}$ & 51.1$^{+8.2}_{-7.1}$ & 10.7$^{+2.3}_{-1.9}$ & 1.7$^{+0.4}_{-0.3}$ & 2.8 & $-1.62$ & $-0.17$ & 45.2 & 3479.1 & 1 & 0.27 & 18.2 \\
\object{SDSS~J084846.11$+$611234.6} & 30.3$^{+6.6}_{-5.5}$ & 8.8$^{+4.1}_{-2.9}$ & 40.0$^{+7.4}_{-6.3}$ & 19.7$^{+4.3}_{-3.6}$ & $<4.3$ & 7.9 & $<-1.64$ & $<+0.15$ & 46.9 & 4280.3 & 4 & 1.11 & 22.0 \\
\object{SDSS~J085116.14$+$424328.8} & 7.0$^{+3.8}_{-2.6}$ & 7.0$^{+3.8}_{-2.6}$ & 13.9$^{+4.8}_{-3.7}$ & 2.4$^{+1.3}_{-0.9}$ & 0.4$^{+0.2}_{-0.1}$ & 3.3 & $-1.90$ & $-0.45$ & 45.2 & 4468.1 & 1 & 0.17 & 29.0 \\
\object{SDSS~J090033.50$+$421547.0} & 81.1$^{+10.0}_{-9.0}$ & 24.7$^{+6.0}_{-4.9}$ & 107.5$^{+11.4}_{-10.3}$ & 20.8$^{+2.6}_{-2.3}$ & 5.2$\pm{0.6}$ & 9.4 & $-1.63$ & $+0.23$ & 47.2 & 3534.0 & 3 & 2.27 & 40.3 \\
\object{SDSS~J091451.42$+$421957.0} & 20.7$^{+5.6}_{-4.5}$ & 7.9$^{+3.9}_{-2.7}$ & 28.5$^{+6.4}_{-5.3}$ & 5.9$^{+1.6}_{-1.3}$ & 0.9$^{+0.3}_{-0.2}$ & 2.6 & $-1.71$ & $-0.25$ & 45.2 & 2945.6 & 1 & 0.38 & 65.1 \\  
\object{SDSS~J093502.52$+$433110.6} & 320.6$^{+18.9}_{-17.9}$ & 159.5$^{+13.7}_{-12.6}$ & 478.8$^{+22.9}_{-21.9}$ & 110.9$^{+6.5}_{-6.2}$ & 16.7$^{+1.0}_{-0.9}$ & 13.1 & $-1.50$ & $+0.06$ & 45.8 & 8467.4 & 1 & 0.09 & 46.2 \\
\object{SDSS~J094202.04$+$042244.5} & 32.7$^{+6.8}_{-5.7}$ & 9.8$^{+4.2}_{-3.1}$ & 43.5$^{+7.7}_{-6.6}$ & 8.0$^{+1.7}_{-1.4}$ & 2.1$\pm{0.4}$ & 6.1 & $-1.71$ & $+0.11$ & 47.1 & 4396.0 & 3 & 1.31 & 35.0 \\
\object{SDSS~J094602.31$+$274407.0} & 4.0$^{+3.2}_{-1.9}$ & $<4.8$ & 5.0$^{+3.4}_{-2.1}$ & 0.8$^{+0.6}_{-0.4}$ & 0.2$\pm{0.1}$ & 7.1 & $-2.16$ & $-0.37$ & 46.9 & 4819.0 & 4 & 0.88 & 8.3 \\
\object{SDSS~J094646.94$+$392719.0} & 13.9$^{+4.8}_{-3.7}$ & 6.8$^{+3.7}_{-2.5}$ & 20.6$^{+5.6}_{-4.5}$ & 0.5$^{+0.2}_{-0.1}$ & 0.1$\pm{0.0}$ & 3.1 & $-2.11$ & $-0.40$ & 46.5 & 4966.0 & 4 & 0.54 & 19.9 \\
\object{SDSS~J095852.19$+$120245.0} & 15.8$^{+5.1}_{-3.9}$ & 2.0$^{+2.6}_{-1.3}$ & 20.8$^{+5.6}_{-4.5}$ & 10.2$^{+3.3}_{-2.5}$ & 2.9$^{+0.9}_{-0.7}$ & 4.4 & $-1.61$ & $+0.19$ & 46.9 & 4505.7 & 4 & 1.01 & 30.8 \\
\object{SDSS~J100054.96$+$262242.4} & 33.2$^{+6.8}_{-5.7}$ & 13.7$^{+4.8}_{-3.6}$ & 47.7$^{+8.0}_{-6.9}$ & 9.5$^{+2.0}_{-1.6}$ & 1.5$\pm{0.3}$ & 2.8 & $-1.64$ & $-0.19$ & 45.2 & 1798.7 & 1 & 1.02 & 13.6 \\
\object{SDSS~J102907.09$+$651024.6} & 114.1$^{+11.7}_{-10.7}$ & 24.6$^{+6.0}_{-4.9}$ & 139.4$^{+12.8}_{-11.8}$ & 10.7$^{+1.1}_{-1.0}$ & 1.9$\pm{0.2}$ & 5.4 & $-1.71$ & $+0.04$ & 46.7 & 4770.0 & 4 & 0.72 & 26.2 \\
\object{SDSS~J103320.65$+$274024.2} & 34.4$^{+6.9}_{-5.8}$ & 24.8$^{+6.1}_{-4.9}$ & 59.2$^{+8.7}_{-7.7}$ & 9.8$^{+2.0}_{-1.7}$ & 1.6$\pm{0.3}$ & 2.6 & $-1.62$ & $-0.17$ & 45.2 & 5077.0 & 1 & 0.13 & 49.6 \\
\object{SDSS~J111119.10$+$133603.8} & 134.5$^{+12.6}_{-11.6}$ & 45.2$^{+7.8}_{-6.7}$ & 179.5$^{+14.4}_{-13.4}$ & 3.1$\pm{0.3}$ & 1.1$\pm{0.1}$ & 7.0 & $-1.84$ & $0.00$ & 47.2 & 6919.0 & 4 & 0.59 & 18.9 \\
\object{SDSS~J111138.66$+$575030.0} & 36.4$^{+7.1}_{-6.0}$ & 9.9$^{+4.3}_{-3.1}$ & 46.4$^{+7.9}_{-6.8}$ & 12.2$^{+2.4}_{-2.0}$ & 1.8$^{+0.4}_{-0.3}$ & 3.4 & $-1.64$ & $-0.19$ & 45.2 & 1676.4 & 1 & 1.18 & 37.6 \\
\object{SDSS~J111830.28$+$402554.0} & 1814.5$^{+43.6}_{-42.6}$ & 543.9$^{+24.3}_{-23.3}$ & 2355.5$^{+49.5}_{-48.5}$ & 92.1$\pm{2.2}$ & 7.8$\pm{0.2}$ & 18.6 & $-1.68$ & $-0.28$ & 44.9 & 4057.0 & 1 & 0.15 & 47.2 \\
\object{SDSS~J111908.67$+$211918.0} & 11630$\pm{150}$ & 987$\pm{13}$ & 12620$\pm{170}$ & 130$\pm{2}$ & 159.5$\pm{2.1}$ & 72.4\tablenotemark{b} & $-1.40$ & $+0.12$ & 45.7 & 2920.0 & 5 & 0.62 & 65.8 \\
\object{SDSS~J111941.12$+$595108.7} & 16.5$^{+5.2}_{-4.0}$ & 8.0$^{+4.0}_{-2.8}$ & 26.4$^{+6.2}_{-5.1}$ & 4.7$^{+1.5}_{-1.1}$ & 0.7$\pm{0.2}$ & 2.7 & $-1.76$ & $-0.32$ & 45.1 & 1501.6 & 1 & 1.33 & 22.4 \\
\object{SDSS~J112224.15$+$031802.6} & 12.6$^{+4.7}_{-3.5}$ & 13.8$^{+4.8}_{-3.7}$ & 26.4$^{+6.2}_{-5.1}$ & 4.3$^{+1.6}_{-1.2}$ & 0.7$^{+0.3}_{-0.2}$ & 3.4 & $-1.80$ & $-0.35$ & 45.2 & 3214.1 & 1 & 0.32 & 22.4 \\
\object{SDSS~J112614.93$+$310146.6} & 87.8$^{+10.4}_{-9.4}$ & 31.5$^{+6.7}_{-5.6}$ & 119.3$^{+12.0}_{-10.9}$ & 25.1$^{+3.0}_{-2.7}$ & 3.9$^{+0.5}_{-0.4}$ & 2.3 & $-1.44$ & $-0.01$ & 45.1 & 4005.0 & 1 & 0.19 & 65.1 \\
\object{SDSS~J113327.78$+$032719.1} & 87.5$^{+10.4}_{-9.3}$ & 28.4$^{+6.4}_{-5.3}$ & 119.2$^{+12.0}_{-10.9}$ & 25.5$^{+3.0}_{-2.7}$ & 4.1$^{+0.5}_{-0.4}$ & 2.9 & $-1.48$ & $-0.02$ & 45.2 & 4062.9 & 1 & 0.20 & 137.3 \\
\object{SDSS~J115954.33$+$201921.1} & $<6.4$ & $<4.8$ & 3.0$^{+2.9}_{-1.6}$ & $<4.1$ & $<1.2$ & 7.6 & $<-1.85$ & $<0.00$ & 47.2 & 6599.0 & 3 & 0.65 & 24.8 \\
\object{SDSS~J123734.47$+$444731.7} & 72.3$^{+9.5}_{-8.5}$ & 28.9$^{+6.4}_{-5.3}$ & 101.1$^{+11.1}_{-10.0}$ & 24.5$^{+3.2}_{-2.9}$ & 3.7$^{+0.5}_{-0.4}$ & 3.9 & $-1.54$ & $-0.08$ & 45.2 & 4257.5 & 1 & 0.18 & 23.1 \\
\object{SDSS~J125415.55$+$480850.6} & 103.7$^{+11.2}_{-10.2}$ & 49.6$^{+8.1}_{-7.0}$ & 154.0$^{+13.4}_{-12.4}$ & 34.0$^{+3.7}_{-3.3}$ & 5.2$^{+0.6}_{-0.5}$ & 3.5 & $-1.47$ & $0.00$ & 45.3 & 3597.7 & 1 & 0.28 & 55.7 \\
\object{SDSS~J131627.84$+$315825.7} & $<3.0$ & 4.0$^{+3.2}_{-1.9}$ & 4.0$^{+3.2}_{-1.9}$ & $<0.9$ & $<0.1$ & 2.6 & $<-2.03$ & $<-0.60$ & 45.1 & 2589.4 & 1 & 0.45 & 44.4 \\
\object{SDSS~J134701.54$+$215401.1} & 69.7$^{+9.4}_{-8.3}$ & 19.6$^{+5.5}_{-4.4}$ & 89.2$^{+10.5}_{-9.4}$ & 19.9$^{+2.7}_{-2.4}$ & 3.1$\pm{0.4}$ & 2.2 & $-1.48$ & $-0.05$ & 45.1 & 2201.1 & 1 & 0.62 & 51.8 \\
\object{SDSS~J135023.68$+$265243.1} & 264.5$^{+17.3}_{-16.3}$ & 133.4$^{+12.6}_{-11.5}$ & 396.7$^{+20.9}_{-19.9}$ & 4.5$\pm{0.3}$ & 1.5$\pm{0.1}$ & 4.3 & $-1.71$ & $-0.02$ & 46.4 & 3813.0 & 6 & 0.81 & 30.8 \\
\object{SDSS~J140331.29$+$462804.8} & 8.0$^{+4.0}_{-2.8}$ & 8.9$^{+4.1}_{-2.9}$ & 17.8$^{+5.3}_{-4.2}$ & 2.7$^{+1.3}_{-0.9}$ & 0.4$^{+0.2}_{-0.1}$ & 3.0 & $-1.87$ & $-0.43$ & 45.1 & 2459.4 & 1 & 0.49 & 73.9 \\
\object{SDSS~J140621.89$+$222346.5} & 1290.0$^{+36.9}_{-35.9}$ & 325.2$^{+19.1}_{-18.0}$ & 1620.8$^{+41.3}_{-40.3}$ & 16.3$\pm{0.5}$ & 11.6$\pm{0.4}$ & 10.8 & $-1.52$ & $-0.25$ & 44.3 & 1524.9 & 1 & 0.58 & 34.1 \\
\object{SDSS~J141028.14$+$135950.2} & 49.2$^{+8.1}_{-7.0}$ & 14.8$^{+4.9}_{-3.8}$ & 63.8$^{+9.0}_{-8.0}$ & 12.2$^{+2.0}_{-1.7}$ & 2.2$^{+0.4}_{-0.3}$ & 4.9 & $-1.67$ & $+0.08$ & 46.7 & 5565.0 & 4 & 0.53 & 31.4 \\
\object{SDSS~J141141.96$+$140233.9} & 46.6$^{+7.9}_{-6.8}$ & 11.8$^{+4.5}_{-3.4}$ & 58.3$^{+8.7}_{-7.6}$ & 13.7$^{+2.3}_{-2.0}$ & 2.3$^{+0.4}_{-0.3}$ & 1.1 & $-1.42$ & $+0.16$ & 45.9 & 3966.0 & 7 & 0.44 & 4.3 \\
\object{SDSS~J141730.92$+$073320.7} & 20.8$^{+5.6}_{-4.5}$ & 4.9$^{+3.4}_{-2.1}$ & 25.6$^{+6.1}_{-5.0}$ & 8.4$^{+2.3}_{-1.8}$ & 1.4$^{+0.4}_{-0.3}$ & 1.6 & $-1.56$ & $+0.05$ & 46.0 & 2784.0 & 7 & 0.99 & 1.4 \\
\object{SDSS~J141949.39$+$060654.0} & 93.1$^{+10.7}_{-9.6}$ & 11.9$^{+4.6}_{-3.4}$ & 105.9$^{+11.3}_{-10.3}$ & 9.4$^{+1.1}_{-1.0}$ & 1.4$\pm{0.2}$ & 3.0 & $-1.66$ & $0.00$ & 46.2 & 5252.0 & 6 & 0.35 & 21.4 \\
\object{SDSS~J141951.84$+$470901.3} & 125.1$^{+12.2}_{-11.2}$ & 28.7$^{+6.4}_{-5.3}$ & 154.4$^{+13.5}_{-12.4}$ & 16.3$^{+1.6}_{-1.5}$ & 2.3$\pm{0.2}$ & 6.2 & $-1.70$ & $+0.07$ & 46.8 & 4816.0 & 4 & 0.79 & 23.2 \\
\object{SDSS~J144741.76$-$020339.1} & 4.9$^{+3.4}_{-2.1}$ & $<4.8$ & 5.9$^{+3.6}_{-2.4}$ & 2.4$^{+1.7}_{-1.1}$ & 0.4$^{+0.3}_{-0.2}$ & 1.7 & $-1.78$ & $-0.20$ & 45.9 & 1923.0 & 7 & 1.87 & 7.7\tablenotemark{c} \\
\object{SDSS~J145334.13$+$311401.4} & 6.0$^{+4.0}_{-2.4}$ & 17.7$^{+5.3}_{-4.2}$ & 23.7$^{+5.9}_{-4.8}$ & 2.0$^{+1.2}_{-0.8}$ & 0.3$^{+0.2}_{-0.1}$ & 3.9 & $-1.96$ & $-0.50$ & 45.3 & 4936.1 & 1 & 0.15 & 51.1 \\
\object{SDSS~J152156.48$+$520238.5} & 41.7$^{+7.5}_{-6.4}$ & 43.0$^{+7.6}_{-6.5}$ & 84.7$^{+10.2}_{-9.2}$ & 1.1$\pm{0.2}$ & 0.2$\pm{0.04}$ & 19.9 & $-2.29$ & $-0.43$ & 47.3 & 5750.0 & 8 & 0.96 & 9.0 \\
\object{SDSS~J152654.61$+$565512.3} & 44.6$^{+7.7}_{-6.6}$ & 22.5$^{+5.8}_{-4.7}$ & 67.0$^{+9.2}_{-8.2}$ & 12.7$^{+2.2}_{-1.9}$ & 1.9$\pm{0.3}$ & 2.7 & $-1.59$ & $-0.15$ & 45.1 & 2690.7 & 1 & 0.41 & 50.1 \\
\object{SDSS~J155837.77$+$081345.8} & 30.6$^{+6.6}_{-5.5}$ & 23.6$^{+5.9}_{-4.8}$ & 54.2$^{+8.4}_{-7.3}$ & 8.9$^{+1.9}_{-1.6}$ & 1.5$\pm{0.3}$ & 2.7 & $-1.64$ & $-0.19$ & 45.2 & 2429.7 & 1 & 0.56 & 40.5 \\
\object{SDSS~J212329.46$-$005052.9} & 548.0$^{+24.4}_{-23.4}$ & 195.4$^{+15.0}_{-14.0}$ & 741.6$^{+28.2}_{-27.2}$ & 13.9$\pm{0.6}$ & 3.9$\pm{0.2}$ & 14.5 & $-1.75$ & $+0.09$ & 47.2 & 4500.0 & 9 & 1.40 & 18.5 \\
\object{SDSS~J230301.45$-$093930.7} & 4.8$^{+3.4}_{-2.1}$ & $<3.0$ & 5.7$^{+3.5}_{-2.3}$ & 3.1$^{+2.2}_{-1.4}$ & 0.9$^{+0.6}_{-0.4}$ & 4.4 & $-1.80$ & $0.00$ & 47.0 & 5887.0 & 3 & 0.66 & 18.7 \\
\object{SDSS~J234145.51$-$004640.5} & 29.6$^{+6.5}_{-5.4}$ & 28.4$^{+6.4}_{-5.3}$ & 58.9$^{+8.7}_{-7.7}$ & 8.6$^{+1.9}_{-1.6}$ & 1.4$\pm{0.3}$ & 2.7 & $-1.65$ & $-0.20$ & 45.2 & 6152.5 & 1 & 0.09 & 62.0 \\
\object{SDSS~J235321.62$-$002840.6} & 81.9$^{+10.1}_{-9.0}$ & 26.6$^{+6.2}_{-5.1}$ & 108.3$^{+11.4}_{-10.4}$ & 28.7$^{+3.5}_{-3.2}$ & 3.1$^{+0.4}_{-0.3}$ & 1.8 & $-1.44$ & $+0.04$ & 45.4 & 3808.4 & 1 & 0.28 & 99.1
\enddata
\tablecomments{\textit{Column (1)} is the SDSS quasar name; \textit{Columns (2) - (4)} give the \xray\ counts in the soft (observed-frame \hbox{$0.5-2$ keV}), hard (observed-frame \hbox{$2-8$ keV}), and full (observed-frame \hbox{$0.5-8$ keV}) bands, respectively; \textit{Column (5)} gives the count rate in the soft band in units of $10^{-3}$\,counts\,s$^{-1}$; \textit{Column (6)} gives the Galactic absorption-corrected flux density at rest-frame $2$ keV in units of $10^{-31}$\,erg\,cm$^{-2}$\,s$^{-1}$\,Hz$^{-1}$, assuming a power-law model with $\Gamma=2.0$; \textit{Column (7)} gives the optical flux density at rest-frame $2500$\,\text{\AA} with units of $10^{-27}$\,erg\,cm$^{-2}$\,s$^{-1}$\,Hz$^{-1}$, obtained from Shen \et (2011) and corrected for Galactic extinction unless otherwise noted; \textit{Column (8)} is the \aox\ parameter; \textit{Column (9)} gives the $\Delta$\aox\ parameter, which is the difference between the measured \aox\ from Column (8) and the predicted \aox , based on the \hbox{\aox-$L_{\nu}(2500\,\text{\AA})$} relation in quasars (given as eq. [3] of Timlin \et 2020); \textit{Column (10)} gives the monochromatic luminosity at a rest-frame wavelength of $5100\,\text{\AA}$ [$\nu L_{\nu}(5100\,\text{\AA})$] computed from the flux densities in Column (7), assuming a UV-optical power-law slope of $\alpha_{\rm uv} = -0.5$ (e.g., Vanden Berk \et 2001); \textit{Columns (11) and (12)} are the archival measurements and respective references for the FWHM of the broad \hb\ line; \textit{Column (13)} is the Eddington ratio, derived using eq. [2] of Shemmer \et (2010); \textit{Column (14)} gives the rest-frame \civ\ equivalent width as described in the text.} 
\tablerefs{Rest-frame optical data obtained from: (1) Shen \et (2011); (2) Mejia-Restrepo \et (2016); (3) Zuo \et (2015); (4) Matthews \et (2021); (5) Boroson \& Green (1992); (6) Shen \& Liu (2012); (7) Plotkin \et (2015); (8) Wu \et (2011); (9) Dix \et (2020).}
\tablenotetext{a}{Errors on the \xray\ counts correspond to the $1\sigma$ level, and were computed according to Tables~1 and 2 of Gehrels (1986) using Poisson statistics. Upper limits were computed according to Kraft \et (1991) and represent the 95\% confidence level; upper limits of $3.0$, $4.8$, and $6.4$ indicate that $0$, $1$, and $2$ \xray\ counts, respectively, have been found within an extraction region of radius $1$\arcsec\ centered on the source's optical position (considering the background within this source-extraction region to be negligible).}
\tablenotetext{b}{Extrapolated from the SDSS spectrum assuming a UV-optical power-law slope of $\alpha_{\rm uv} = -0.5$ (e.g., Vanden Berk \et 2001) and corrected for Galactic extinction using the $A_{\rm V}$ value from Schlegel \et (1998) and the extinction curve from Cardelli \et (1989).}
\tablenotetext{c}{Taken from Plotkin \et (2015).}
\end{deluxetable*}
\end{longrotatetable}

\section{Results and Discussion}
\label{sec:results}
Our goal is to test whether \xray\ data can strengthen current optical-UV indicators of \lledd\ such as those provided by the \civ\ spectroscopic parameter space. The following provides a step-by-step description of the analyses performed to test this hypothesis.

\begin{figure*}
\plotone{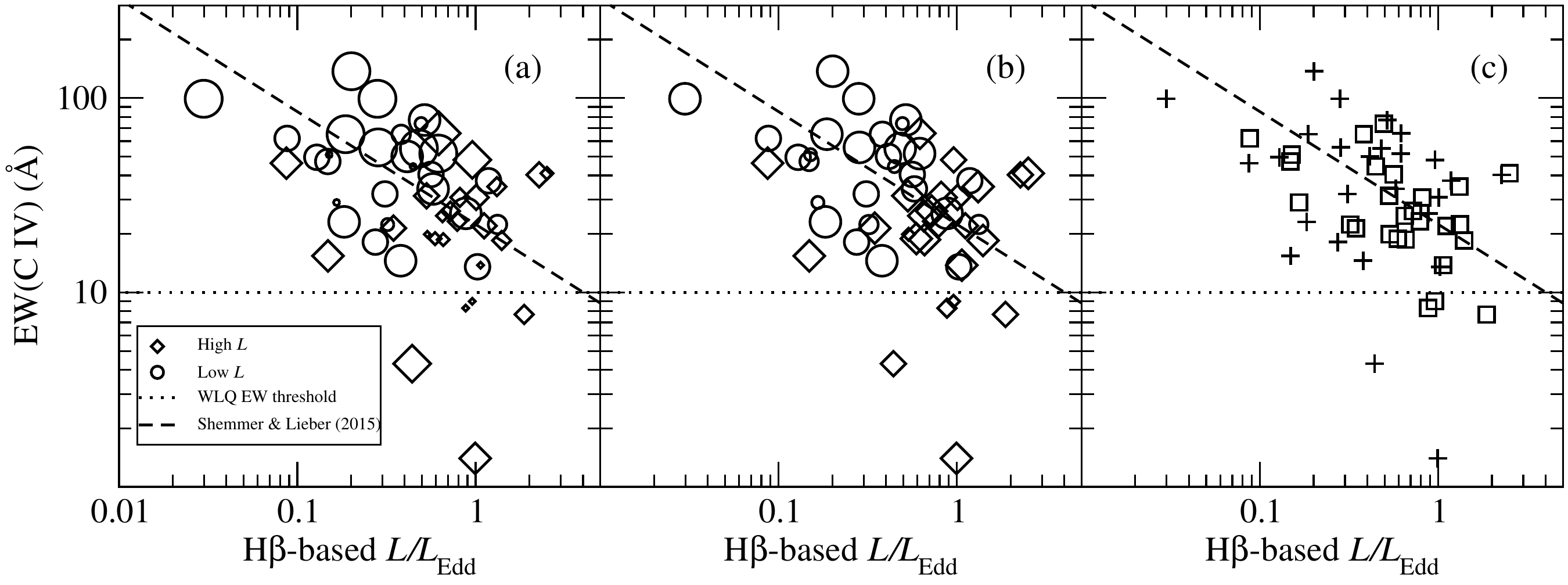}
\caption{The EW(\civ) vs. \lledd\ anti-correlation. For panels (a) and (b), circles represent the Low $L$ sources \hbox{($\log L_{2500\,\text{\AA}} < 30.7$ erg s$^{-1}$ Hz$^{-1}$)} and diamonds represent the High $L$ sources. Larger symbol sizes correspond to sources with larger (i.e., less negative) \aox\ values in (a). In (b), symbol sizes denote the $\sigma$ values of \daox, with the largest symbols corresponding to \daox\ values within 1$\sigma$ ($\pm{0.131}$; e.g., Just \et 2007). In (c), squares (pluses) represent objects with an \aox\ value below (above) the median value of $-1.64$. The dotted lines show the $10\,\text{\AA}$ EW threshold for defining WLQs (Diamond-Stanic \et 2009), and the dashed lines show the regression lines corresponding to Equation [2] of Shemmer \& Lieber (2015). It appears that the additional \xray\ information does not contribute any new trends to the data.}
\label{fig:civ_lledd}
\end{figure*}

Figure~\ref{fig:civ_lledd} shows a significant anti-correlation between EW(\civ) and \hb-based \lledd\ for our sources, with a Spearman-rank correlation coefficient ($r_{s}$) of $-0.42$ and a chance probability ($p$) of $1.8\times10^{-3}$, confirming the results of Shemmer \& Lieber (2015). We find that two of our sources deviate significantly ($\gtrsim4\sigma$) from this correlation; these are SDSS J1411$+$1402 and SDSS J1417$+$0733, which are WLQs with EW(\civ) values of $4.3\,\text{\AA}$ and $1.4\,\text{\AA}$, respectively. However, the exclusion of these sources does not significantly impact the correlation \hbox{($r_{s} = -0.43$, $p = 1.7\times10^{-3}$).} To test whether \xray\ information can minimize the scatter in this correlation, symbol sizes in Figure~\ref{fig:civ_lledd} (a) depend on the objects' \aox\ values, and symbol sizes in Figure~\ref{fig:civ_lledd} (b) depend on \daox. We do not find any trends stemming from this sorting by \aox\ or \daox.
We also ran partial correlations distinguishing between \xray\ strong (weak) sources if these are above (below) the median \aox\ for our sample which is $-1.64$; these correlation coefficients and chance probabilities are shown in Table~\ref{tab:corr}. We find a stronger anti-correlation for the \xray\ weak sources ($r_{s} = -0.55$, $p = 4.0\times10^{-3}$) than the \xray\ strong sources ($r_{s} = -0.26$, $p = 0.18$), which can be seen in Figure~\ref{fig:civ_lledd} (c).

\begin{figure*}
\plotone{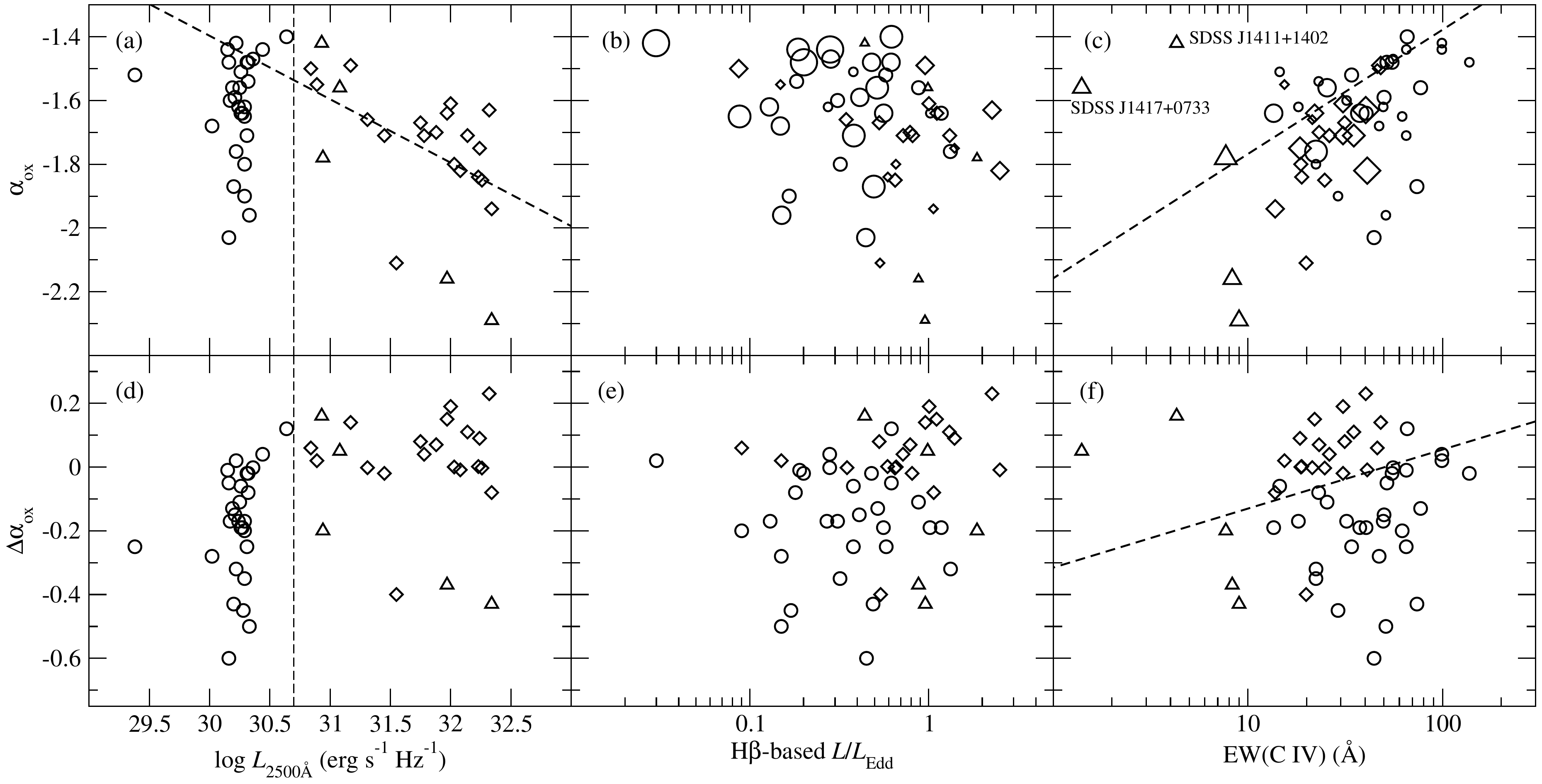}
\caption{Optical-to-X-ray spectral slope (\aox) (top) and \daox\ (bottom) vs $L_{2500\,\text{\AA}}$ (left), \hb-based \lledd\ (center), and EW(\civ) (right). Circles represent the Low $L$ sources, diamonds represent the High $L$ sources, and triangles represent WLQs (all of which are included in the High $L$ sub-sample); the threshold between low and high luminosity is shown by the vertical dashed line in panels (a) and (d). The bold dashed lines in panels (a), (c), and (f) show the best fit relations found in Timlin \et (2020). Symbol sizes in panels (b) and (c) denote the EW(\civ) and \lledd\ values, respectively, with larger symbols corresponding to larger values. Our High $L$ sources appear to follow the \aox-$L_{2500\,\text{\AA}}$ relation of Timlin \et (2020), however our Low $L$ sources seem to spoil the relation, presumably given the selection bias inherent in the majority of these sources as explained in the text.}
\label{fig:aox}
\end{figure*}

To investigate whether our sample is biased with respect to the \aox\ diagnostic, in Figure~\ref{fig:aox} we show \aox\ and \daox\ vs $L_{2500\,\text{\AA}}$ (left), \hb-based \lledd\  (center), and EW(\civ) (right).
In spite of the fact that no significant evolution in the \xray\ properties of quasars has been observed across cosmic time (e.g., Shemmer \et 2005; Nanni \et 2017; Vito \et 2019), we also ran partial correlations after sorting our sample by luminosity, creating ``Low $L$" and ``High $L$" subsets with Low $L$ corresponding to the low redshift \hbox{($z < 0.8$)} sources and defined as \hbox{$\log L_{2500\,\text{\AA}} < 30.7$ erg s$^{-1}$ Hz$^{-1}$.}
Spearman-rank correlation coefficients and chance probabilities for the full sample and each sub-sample are also presented in Table~\ref{tab:corr}.

\newpage
\startlongtable
\begin{deluxetable}{llcc}
\tabletypesize{\footnotesize}
\tablecolumns{4}
\tablecaption{\aox (\daox) Spearman-rank Partial Correlations\label{tab:corr}}
\tablehead{ 
\colhead{vs.} &
\colhead{Control} &
\colhead{$r_{s}$} &
\colhead{$p$} \\
\colhead{(1)} &
\colhead{(2)} &
\colhead{(3)} &
\colhead{(4)}
}
\startdata
\multicolumn{4}{c}{Full Sample (53 sources)} \\
\noalign{\smallskip}\hline\noalign{\smallskip}
$L_{2500\,\text{\AA}}$ & {} & $-0.40$ ($+0.50$) & \textbf{0.003} \textbf{($\mathbf{1.3\times10^{-4}}$)} \\
$L_{2500\,\text{\AA}}$ & \lledd & $-0.27$ ($+0.47$) & \textbf{0.05 ($\mathbf{5.1\times10^{-4}}$)} \\
$L_{2500\,\text{\AA}}$ & EW(\civ) & $-0.28$ ($+0.54$) & \textbf{0.04 ($<0.0001$)} \\
\noalign{\smallskip}\hline
\lledd & {} & $-0.34$ ($+0.22$) & \textbf{0.01} ($>0.05$) \\
\lledd & $L_{2500\,\text{\AA}}$ & $-0.18$ ($-0.05$) & $>0.05$ ($>0.05$) \\
\lledd & EW(\civ) & $-0.23$ ($+0.22$) & $>0.05$ ($>0.05$) \\
\noalign{\smallskip}\hline
EW(\civ) & {} & $+0.36$ ($-0.05$) & \textbf{0.008} ($>0.05$) \\
EW(\civ) & $L_{2500\,\text{\AA}}$ & $+0.23$ ($+0.22$) & $>0.05$ ($>0.05$) \\
EW(\civ) & \lledd & $+0.26$ ($+0.04$) & $>0.05$ ($>0.05$) \\
\noalign{\smallskip}\hline
\multicolumn{4}{c}{Low $L$ (29 sources)} \\
\noalign{\smallskip}\hline\noalign{\smallskip}
$L_{2500\,\text{\AA}}$ & {} & $+0.12$ ($+0.25$) & $>0.05$ ($>0.05$) \\
$L_{2500\,\text{\AA}}$ & \lledd & $+0.12$ ($+0.24$) & $>0.05$ ($>0.05$) \\
$L_{2500\,\text{\AA}}$ & EW(\civ) & $+0.06$ ($+0.21$) & $>0.05$ ($>0.05$) \\
\noalign{\smallskip}\hline
\lledd & {} & $-0.02$ ($-0.06$) & $>0.05$ ($>0.05$) \\
\lledd & $L_{2500\,\text{\AA}}$ & $+0.004$ ($-0.01$) & $>0.05$ ($>0.05$) \\
\lledd & EW(\civ) & $+0.10$ ($+0.06$) & $>0.05$ ($>0.05$) \\
\noalign{\smallskip}\hline
EW(\civ) & {} & $+0.40$ ($+0.41$) & \textbf{0.03 (0.03)} \\
EW(\civ) & $L_{2500\,\text{\AA}}$ & $+0.39$ ($+0.39$) & \textbf{0.04 (0.04)} \\
EW(\civ) & \lledd & $+0.41$ ($+0.41$) & \textbf{0.03 (0.03)} \\
\noalign{\smallskip}\hline
\multicolumn{4}{c}{High $L$ (24 sources)} \\
\noalign{\smallskip}\hline\noalign{\smallskip}
$L_{2500\,\text{\AA}}$ & {} & $-0.62$ ($-0.09$) & \textbf{0.001} ($>0.05$) \\
$L_{2500\,\text{\AA}}$ & \lledd & $-0.61$ ($-0.17$) & \textbf{0.002} ($>0.05$) \\
$L_{2500\,\text{\AA}}$ & EW(\civ) & $-0.66$ ($-0.13$) & \textbf{$\mathbf{5.9\times10^{-4}}$} ($>0.05$) \\
\noalign{\smallskip}\hline
\lledd & {} & $-0.20$ ($+0.12$) & $>0.05$ ($>0.05$) \\
\lledd & $L_{2500\,\text{\AA}}$ & $+0.16$ ($+0.18$) & $>0.05$ ($>0.05$) \\
\lledd & EW(\civ) & $-0.22$ ($+0.09$) & $>0.05$ ($>0.05$) \\
\noalign{\smallskip}\hline
EW(\civ) & {} & $+0.27$ ($+0.39$) & $>0.05$ ($>0.05$) \\
EW(\civ) & $L_{2500\,\text{\AA}}$ & $+0.39$ ($+0.40$) & $>0.05$ ($>0.05$) \\
EW(\civ) & \lledd & $+0.29$ ($+0.39$) & $>0.05$ ($>0.05$) \\
\noalign{\smallskip}\hline
\multicolumn{4}{c}{\xray\ Weak (26 sources)} \\
\noalign{\smallskip}\hline\noalign{\smallskip}
$L_{2500\,\text{\AA}}$ & {} & $-0.27$ ($+0.54$) & $>0.05$ \textbf{(0.005)} \\
$L_{2500\,\text{\AA}}$ & \lledd & $-0.27$ ($+0.42$) & $>0.05$ \textbf{(0.04)} \\
$L_{2500\,\text{\AA}}$ & EW(\civ) & $-0.20$ ($+0.54$) & $>0.05$ \textbf{(0.006)} \\
\noalign{\smallskip}\hline
\lledd & {} & $-0.08$ ($+0.39$) & $>0.05$ \textbf{(0.05)} \\
\lledd & $L_{2500\,\text{\AA}}$ & $+0.08$ ($+0.15$) & $>0.05$ ($>0.05$) \\
\lledd & EW(\civ) & $+0.05$ ($+0.38$) & $>0.05$ ($>0.05$) \\
\noalign{\smallskip}\hline
EW(\civ) & {} & $+0.21$ ($-0.15$) & $>0.05$ ($>0.05$) \\
EW(\civ) & $L_{2500\,\text{\AA}}$ & $+0.10$ ($+0.15$) & $>0.05$ ($>0.05$) \\
EW(\civ) & \lledd & $+0.20$ ($+0.09$) & $>0.05$ ($>0.05$) \\
\noalign{\smallskip}\hline
\multicolumn{4}{c}{\xray\ Strong (27 sources)} \\
\noalign{\smallskip}\hline\noalign{\smallskip}
$L_{2500\,\text{\AA}}$ & {} & $-0.07$ ($+0.79$) & $>0.05$ $\mathbf{(<0.0001)}$ \\
$L_{2500\,\text{\AA}}$ & \lledd & $+0.05$ ($+0.78$) & $>0.05$ $\mathbf{(<0.0001)}$ \\
$L_{2500\,\text{\AA}}$ & EW(\civ) & $+0.07$ ($+0.81$) & $>0.05$ $\mathbf{(<0.0001)}$ \\
\noalign{\smallskip}\hline
\lledd & {} & $-0.39$ ($+0.17$) & \textbf{0.04} ($>0.05$) \\
\lledd & $L_{2500\,\text{\AA}}$ & $-0.39$ ($-0.09$) & \textbf{0.05} ($>0.05$) \\
\lledd & EW(\civ) & $-0.32$ ($+0.17$) & $>0.05$ ($>0.05$) \\
\noalign{\smallskip}\hline
EW(\civ) & {} & $+0.45$ ($-0.02$) & \textbf{0.02} ($>0.05$) \\
EW(\civ) & $L_{2500\,\text{\AA}}$ & $+0.45$ ($+0.33$) & \textbf{0.02} ($>0.05$) \\
EW(\civ) & \lledd & $+0.39$ ($+0.03$) & \textbf{0.05} ($>0.05$) \\
\noalign{\smallskip}\hline
\enddata
\tablecomments{\textit{Column (1)} is the parameter that was correlated with \aox\ (\daox); \textit{Column (2)} is the controlled parameter - If entry is empty, we only calculated the correlation between \aox\ (\daox) and the parameter in the first column; otherwise, we calculated the partial correlation between \aox\ (\daox) and the parameter in the first column while controlling for the parameter in the second column; \textit{Columns (3) and (4)} are the Spearman-rank correlation coefficient and chance probability, respectively.
Significant correlations are shown in bold and defined as \hbox{$p < 0.05$}. Low $L$ (High $L$) corresponds to objects with \hbox{$\log L_{2500\,\text{\AA}}$ below (above) $30.7$ erg s$^{-1}$ Hz$^{-1}$}. \xray\ weak (strong) corresponds to objects with an \aox\ value below (above) the median value of $-1.64$.} 
\end{deluxetable}

We find a significant anti-correlation between \aox\ and $L_{2500\,\text{\AA}}$ consistent with previous studies (e.g., Timlin \et 2020), which becomes stronger with the exclusion of our $29$ Low $L$ objects (see Table~\ref{tab:corr}). This deviation may be due to the \aox-$L_{2500\,\text{\AA}}$ relation breaking down at some low $L_{2500\,\text{\AA}}$ value, which could also lead to the very low \daox\ values in some of our low luminosity objects (see Figure~\ref{fig:aox}). The additional exclusion of the \xray\ weak, high-luminosity source SDSS J0946$+$3927 as well as WLQs SDSS J0946$+$2744 and SDSS J1521$+$5202, gives an even stronger correlation with $r_{s} = -0.68$ and $p = 6.3\times10^{-4}$ (see Figure~\ref{fig:aox}). 

One notable finding in Table~\ref{tab:corr} is the apparent strong \daox-$L_{2500\,\text{\AA}}$ correlation for the full sample as well as the \xray\ strong sources. The corresponding significantly smaller correlation coefficients found in the High $L$/ Low $L$ and \xray\ weak sub-samples, as well as visual inspection of Figure~\ref{fig:aox}, suggest that such a strong correlation is not to be expected. Direct comparison with other samples (e.g., Steffen \et 2006; Just \et 2007) confirms that our results for the High $L$/Low $L$ and \xray\ weak sources are consistent with the trends found in previous studies. Therefore, the strong \daox-$L_{2500\,\text{\AA}}$ correlation may be due to a small sample bias as well as the significant scatter (or the breakdown of the \aox-$L_{2500\,\text{\AA}}$ correlation) at low luminosity (see Figure~\ref{fig:aox} (a) and (d)), which is related to a bias in favor of low-luminosity sources and the way they were selected as discussed further below. 

The increased scatter at low luminosity may also have a contribution from larger amplitudes of \xray\ variability which, together with the fact that the \xray\ and optical-UV measurements are non-contemporaneous, would produce uncertainties on the order of \daox$\gtrsim 0.1$ (e.g., Vagnetti \et 2010, 2013). To investigate the potential effects of variability further, we examined the \chandra\ observations of the eight sources that have $>300$ counts in the full band (i.e., non-``snapshots"; see Section~\ref{sec:obs}, Figure~\ref{fig:hist}, and Table~\ref{tab:uv-opt}). We split each of these observations into two sub-exposures with equal exposure times and compared the count rates between the two sub-exposures; the rest-frame time difference between each pair of sub-exposures is in the range $\sim1-10$ hr. We found that in all cases the count rates in both sub-exposures were consistent with each other, within the errors, indicating the absence of short-term variability. 

For the full sample and High $L$ sub-sample, we find that the correlations between \aox\ and \lledd\ are weaker than the respective \aox-$L_{2500\,\text{\AA}}$ correlations (see Table~\ref{tab:corr}). These weaker correlations may be due to the inherent dependence of \lledd\ on $L_{2500\,\text{\AA}}$ and black-hole mass and the additional uncertainties associated with estimating \lledd\ (see, e.g., Shemmer \et 2008). To see if the EW(\civ) parameter contributes to this anti-correlation, Figure~\ref{fig:aox} (b) shows larger symbols that correspond to larger values of EW(\civ); however no trend with EW(\civ) has been found.
 
We find that the correlation between \aox\ and EW(\civ) for the full sample is stronger than the corresponding \aox-\lledd\ correlation, yet not as strong as the \aox-$L_{2500\,\text{\AA}}$ correlation (see Table~\ref{tab:corr}). As can be seen in Figure~\ref{fig:aox}, there are two sources that are significant outliers; these are the WLQs SDSS J1411$+$1402 and SDSS J1417$+$0733. Comparison with Luo \et (2015) shows the same \aox\ values as those calculated in this work, and exclusion of these sources significantly improves the correlation to $r_{s} = 0.41$ and \hbox{$p = 4.3\times10^{-3}$.} The \aox-EW(\civ) correlation also seems to hold in almost any subsample (see Table~\ref{tab:corr}), notwithstanding effects due to changing sample size, which supports the results of Timlin \et (2021) that \aox\ is expected to be correlated with EW(\civ) as an indicator of the shape of the spectral energy distribution. To see if the \lledd\ parameter contributes to this correlation, Figure~\ref{fig:aox} (c) shows larger symbols that correspond to larger values of \lledd; however no trend with \lledd\ has been found.

Overall, Table~\ref{tab:corr} shows a significant difference between the Low $L$ and High $L$ sub-samples with respect to the \aox\ (\daox) vs. $L_{2500\,\text{\AA}}$ and \lledd\ correlations. The Low $L$ sources do not exhibit significant correlations between $L_{2500\,\text{\AA}}$ or \lledd\ and \aox\ and, therefore, spoil the correlations for the entire sample. This difference may be a consequence of the fact that most of these sources were originally selected to have a small range in $L_{2500\,\text{\AA}}$, but a large range in the accretion-rate diagnostics FWHM(\hb) and R(Fe~{\sc ii})\footnote{Defined as the ratio of the equivalent width of the Fe~{\sc ii} emission-line complex in the rest-frame wavelength range $4344 - 4684\,\text{\AA}$ and the equivalent width of broad \hb.} (see Rivera \et 2022, under review), which likely contribute to the large range in \aox.

The \xray\ weak sub-sample exhibits the same trends as the High $L$ sub-sample, namely that the \aox-$L_{2500\,\text{\AA}}$ correlations are the strongest, and the \aox-\lledd\ correlations are the weakest; while the \xray\ strong sub-sample shares one trend with the Low $L$ sub-sample, in which the \aox-EW(\civ) correlations are the strongest.

To quantify the potential contribution of the \aox\ parameter to the Eddington ratio estimate, a multiple regression analysis was performed using the \lledd\ values from Table~\ref{tab:uv-opt} as the dependent variable, and the \aox, $L_{2500\,\text{\AA}}$, and EW(\civ) values from Table~\ref{tab:uv-opt} as the independent variables. These regressions include combinations of the above parameters with linear, interaction, and quadratic terms; each with and without an intercept. The results of these regressions suggest that \aox\ does not contribute significantly to creating a diagnostic to \lledd\ for our entire sample. The linear model with the best fit has the form:

\begin{equation}
\begin{gathered}
\small
L/L_{\rm Edd} = \alpha + \beta\,\text{\aox} + \gamma\,\text{{\rm log}~$L_{2500\,\text{\AA}}$} + \delta\,\text{EW(C~{\sc iv})}
\end{gathered}
\end{equation}
where
\begin{equation*}
\begin{gathered}
\alpha = -9.0\pm{2.6} \\
\beta = 0.1\pm{0.4} \\
\gamma = 0.3\pm{0.1} \\
\delta = -2.8\times10^{-3}\pm{2.8\times10^{-3}}.
\end{gathered}
\end{equation*}
 
We note that $\beta$ and $\delta$ are consistent with zero, suggesting that only $L_{2500\,\text{\AA}}$ contributes to \lledd, which may be a result of our small sample size; we therefore cannot identify a linear combination of these observables that gives us a meaningful \lledd\ indicator. A similar analysis using only the \xray\ weak sources yields similar results, with still no significant contribution from \aox\ to the \lledd\ parameter.

\section{Summary}
\label{sec:summary}
We present correlations between \aox, \daox, $L_{2500\,\text{\AA}}$, \hb-based \lledd, and EW(\civ) in the search for a robust \lledd\ estimate. Our analysis, based on a sample of $53$ radio-quiet quasars without broad absorption lines, yields consistent results with previous studies when it comes to the EW(\civ)-\lledd\ and \aox-EW(\civ) relations. We also find a strong anti-correlation between \aox\ and $L_{2500\,\text{\AA}}$ for sources with lower \aox\ values (and high luminosity) that is not present for the sources with higher \aox\ values (and low luminosity), which is most likely due to a selection bias among our low luminosity sources. However, our results for the full sample do not show that \aox\ significantly improves the strong EW(\civ)-\lledd\ anti-correlation. 

A larger sample size of preferentially high-luminosity, high-redshift sources (whereby UV data can be obtained from optical spectra, e.g., from SDSS) is needed for testing the correlations presented in this work in an unbiased way and drawing firmer conclusions. Since luminous sources that have high-quality archival \xray, \civ, and \hb\ measurements are rare, we plan to obtain \xray\ snapshot observations of sources from the largest, uniform compilation of high-redshift quasars with \hb\ measurements (Matthews \et 2021), thus more than doubling the current inventory. In future investigations, we plan to measure the \civ\ blueshift (using the [O~{\sc iii}]-based systemic redshift) and compute the \civ\ ``distance" as proposed by Rivera \et (2020), to replace the use of the \civ\ EW and blueshift separately. Furthermore, we plan to apply corrections to \hb-based Eddington ratios based on R(Fe~{\sc ii}) measurements (e.g., Du \& Wang 2019), and include $\Gamma$-based Eddington ratio estimates for high-redshift quasars having deep \xray\ observations (i.e., Shemmer \et 2008) which could potentially come from future \xray\ missions, e.g., Athena (Nandra \et 2013). 

Our pilot investigation, based on an archival sample includes all three basic \lledd\ ingredients, and, as the first of its kind, will pave the way for larger, more systematic investigations of these parameters to identify the most reliable Eddington ratio indicator for quasars.

\acknowledgements
We gratefully acknowledge \chandra\ grants \hbox{AR8--19014X} (A.~M., O.~S.) and G07--18110X (G.~T.~R.). We thank an anonymous referee for providing valuable feedback that helped improve this manuscript. This research has made use of the NASA/IPAC Extragalactic Database (NED) which is operated by the Jet Propulsion Laboratory, California Institute of Technology, under contract with the National Aeronautics and Space Administration, as well as NASA's Astrophysics Data System Bibliographic Services. This research has also made use of the data provided by the High Energy Astrophysics Science Archive Research Center (HEASARC), which is a service of the Astrophysics Science Division at NASA/GSFC and the High Energy Astrophysics Division of the Smithsonian Astrophysical Observatory. 


\end{document}